\documentclass[aps,prl,twocolumn,showpacs,amsmath,amssymb,superscriptaddress]{revtex4}
\usepackage{graphicx}
\usepackage{bm}

\newcommand{\ket}[1]{|#1\rangle}
\newcommand{\braket}[1]{\langle #1 \rangle}

\def\dd{\mathrm{d}}
\def\ee{\mathrm{e}}
\def\ii{\mathrm{i}}
\def\vnabla{\bm{\nabla}}

\def\rot{\vnabla\times}

\def\dgg#1{\{#1\}^{\dagger}}

\def\Hc{\mathrm{H.c.}}

\def\Im{\mathrm{Im}}

\def\half{\frac{1}{2}}

\def\diez{\varepsilon_0}
\def\dieb{\varepsilon_{\mathrm{bg}}}
\def\muz{\mu_0}

\def\wex{\omega}
\def\wbx{\varOmega}
\def\wres{\varOmega}
\def\wT{\omega_{\mathrm{T}}}
\def\DLT{\varDelta_{\mathrm{LT}}}
\def\Dbx{\varDelta_{\mathrm{bx}}}

\def\mex{m_{\mathrm{ex}}}

\def\mbx{m_{\mathrm{bx}}}

\def\dampex{\gamma_{\mathrm{ex}}}
\def\dampbx{\gamma_{\mathrm{bx}}}

\def\ccxr{g}

\def\thick{d}

\def\win{\omega_{\mathrm{in}}}

\def\ketg{\ket{\mathrm{g}}}

\def\css{S}

\def\mcss{\mathbf{S}}

\def\crd{W}

\def\mcrd{\mathbf{W}}

\def\wvfbx{F}

\def\rmfbx{\varPhi}

\def\cmfbx{G}

\def\cmfex{g}

\def\eV{\mathrm{eV}}
\def\meV{\mathrm{meV}}
\def\microeV{\mu\mathrm{eV}}

\def\microm{\mu\mathrm{m}}
\def\nm{\mathrm{nm}}
\def\Watt{\mathrm{W}}
\def\ps{\mathrm{ps}}

\def\eV{\mathrm{eV}}
\def\meV{\mathrm{meV}}
\def\microeV{\mu\mathrm{eV}}

\def\microm{\mu\mathrm{m}}
\def\nm{\mathrm{nm}}
\def\Watt{\mathrm{W}}
\def\ps{\mathrm{ps}}
\def\ns{\mathrm{ns}}
\def\sec{\mathrm{s}}

\def\vunit{\bm{e}}

\def\vr{\bm{r}}

\def\vdimP{\bm{\mathcal{P}}}

\def\dimP{M}
\def\vdimP{\bm{\mathcal{P}}}

\def\munit{\mathbf{1}}

\def\mG{\bm{\mathsf{G}}}

\def\oH{H}
\def\oHem{H_{\mathrm{em}}}
\def\oHint{H_{\mathrm{int}}}

\def\oHex{H_{\mathrm{ex}}}

\def\oHres{H_{\mathrm{res}}}

\def\oex{b}
\def\oexd{b^{\dagger}}

\def\obx{B}
\def\obxd{B^{\dagger}}
\def\ores{d}
\def\oresd{d^{\dagger}}

\def\ovE{\bm{E}}

\def\ovPex{\bm{P}_{\mathrm{ex}}}

\def\hex{\hat{b}}

\def\hexone{\hat{b}^{(1)}}

\def\hbx{\hat{B}}

\def\hvPex{\hat{\bm{P}}_{\mathrm{ex}}}

\def\hvJz{\hat{\bm{J}}_0}
\def\hvE{\hat{\bm{E}}}
\def\hvEz{\hat{\bm{E}}_0}

\def\hrsrc{\hat{\mathcal{D}}}

\begin{document}

\title{Entangled-photon generation in nano-to-bulk crossover regime}

\author{Motoaki Bamba}
\altaffiliation[Present address: ]{Laboratoire Mat\'eriaux et Ph\'enom\`enes Quantiques, Universit\'e Paris Diderot-Paris 7 et CNRS, 
Case 7021, B\^atiment Condorcet, 75205 Paris Cedex 13, France.
E-mail: motoaki.bamba@univ-paris-diderot.fr}
\affiliation{Department of Materials Engineering Science, 
Osaka University, Toyonaka, Osaka 560-8631, Japan}
\author{Hajime Ishihara}

\affiliation{Department of Physics and Electronics, 
Osaka Prefecture University, Sakai, Osaka 599-8631, Japan}

\date{\today}

\begin{abstract}
We have theoretically investigated a generation 
of entangled photons from biexcitons in a semiconductor film
with thickness in nano-to-bulk crossover regime.
In contrast to the cases of quantum dots and bulk materials,
we can highly control 
the generated state of entangled photons
by the design of peculiar energy structure of exciton-photon coupled modes
in the thickness range between nanometers and micrometers.
Owing to the enhancement of radiative decay rate of excitons
at this thickness range,
the statistical accuracy of generated photon pairs
can be increased
beyond the trade-off problem with the signal intensity.
Implementing an optical cavity structure,
the generation efficiency can be enhanced 
with keeping the high statistical accuracy.
\end{abstract}

\pacs{42.65.Lm, 42.50.Nn, 71.35.-y, 71.36.+c}
\keywords{}

\maketitle

Entangled photons play an important role in quantum information technologies, and the pursuit of high-quality generation of them has become an active research area
in the fields of quantum optics and solid-state physics. In addition to the standard 
method of generating the entangled-photon pairs by using the parametric down-conversion (PDC) in second order nonlinear crystals \cite{Kwiat}, 
the generation scheme using a semiconductor quantum dot \cite{Akopian,Stevenson} attracts much attention because we can generate a pure single pair of entangled photons that is highly desired in quantum information processing. On the other hand, the development of entangled photons as an excitation light source is becoming of growing importance for the next-generation technologies of fabrication and chemical reaction \cite{Sasaki}. For this purpose,
extra high-power and high-quality entangled-photon beams are absolutely necessary. 
However, any schemes that lead to the generation of such photon beams have not been found yet,
although several generation schemes have been investigated so far.

\begin{figure}[tbp] 
\includegraphics[width=\hsize]{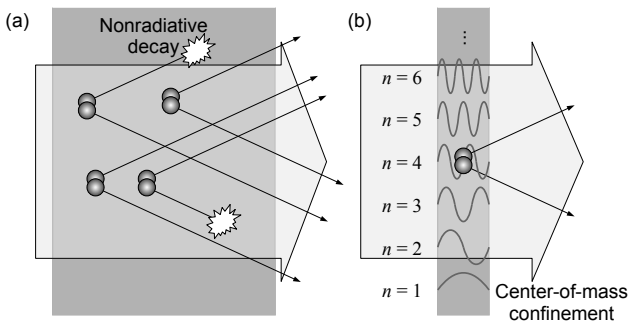}
\caption{Sketches of entangled-photon generation from biexcitons (RHPS).
(a) In a bulk semiconductor, 
statistical accuracy worsens by the nonradiative decay
of generated photons. 
(b) In a nano-film, most of photons can go outside by the exciton superradiance,
and better accuracy is expected. Center-of-mass motions
of excitons and biexcitons are confined in the film.}
\label{fig:1}
\end{figure}
For the high-power generation of entangled photons, one of the promising candidates is the method involving the biexciton-resonant hyperparametric scattering (RHPS) demonstrated in a CuCl bulk crystal \cite{Eda,Oohata}. This generation method is highly efficient owing to the giant two-photon absorption in CuCl crystals \cite{Itoh,Biexciton}. However, as indicated in the experiments \cite{Eda,Oohata}, 
we must consider not only the generation efficiency but also the statistical accuracy,
i.e., accidentally emitted photon pairs (noise),
which come from independent biexcitons and have no entanglement (see Fig.~1(a)).
In the scheme for high-power generation of the entangled photons, 
the amount of entangled pairs (signal) and signal-to-noise ($S/N$) ratio
are always a trade-off,
because the nonradiative damping of excitons
significantly worsens the $S/N$ ratio of the entangled-photon generation.
Although this trade-off problem is seemingly inevitable
by the exsistence of the nonradiative damping
under the resonant excitation of the elementary excitation in condensed matters,
we have revealed that
it can be overcome by the enhancement of radiative decay rate of excitons
in the crossover regime between the quantum dots and bulk materials
\cite{Ichimiya}.
Further, we can also control the characteristics of generated photons
by a variety of degrees of freedom in condensed matters
and by the peculiar energy structure in the crossover regime
\cite{IKS,Shouji,Ichimiya}.

Our calculation method is based on the quantum electrodynamics (QED)
theory for excitons \cite{newQED}.
Compared to the previous theories \cite{Savasta}, we have developed a  
framework applicable to nanometer and submicrometer films including
explicit degrees of freedom of the center-of-mass motion of confined excitons.
We consider excitons as pure bosons
and introduce an exciton-exciton interaction into the system
to discuss the RHPS process.
The Hamiltonian of the excitonic system is represented as
\begin{equation} \label{eq:oHex} 
\oHex = \sum_{\mu} \hbar\wex_{\mu} \oexd_{\mu} \oex_{\mu}
+ \frac{1}{2} \sum_{\mu,\nu,\mu',\nu'}
V_{\mu,\nu;\mu',\nu'} \oexd_{\mu} \oexd_{\nu} \oex_{\nu'} \oex_{\mu'},
\end{equation}
where $\oex_{\mu}$ is the annihilation operator of an exciton in state $\mu$,
$\wex_{\mu}$ is its eigen\-frequency,
and $V_{\mu,\nu;\mu',\nu'}$ is the nonlinear coefficient.
The equation of excitons' motion
is derived in $\omega$ (frequency) domain as \cite{note}
\begin{align} \label{eq:motion-exciton} 
& [\hbar\wex_{\mu}-\hbar\omega-\ii\dampex/2] \hex_{\mu}(\omega)
\nonumber \\ &
= \int\dd\vr\ \vdimP_{\mu}^*(\vr) \cdot \hvE^+(\vr,\omega)
  + \hat{\cal D}_{\mu}(\omega)
\nonumber \\ &
- \sum_{\nu,\mu',\nu'} V_{\mu,\nu;\mu',\nu'}
  \int_{-\infty}^{\infty}\dd t\ \frac{\ee^{\ii\omega t}}{2\pi}
  \oexd_{\nu}(t) \oex_{\nu'}(t) \oex_{\mu'}(t).
\end{align}
Here, $\hvE^+(\vr,\omega)$ is the positive-frequency Fourier transform
of the electric field at position $\vr$.
$\vdimP_{\mu}(\vr)$ is the coefficient of excitonic polarization
$\ovPex(\vr) = \sum_{\mu} \vdimP_{\mu}(\vr) \oex_{\mu} + \Hc$,
and is represented as $\vdimP_{\mu}(\vr) = {\bf M} g_{\mu}(\vr)$
with the transition dipole moment ${\bf M}$
and the exciton center-of-mass wavefunction $g_{\mu}(\vr)$.
$\hat{\cal D}_{\mu}(\omega)$ is the fluctuation operator 
due to the excitonic nonradiative damping with rate $\dampex$.
The equation of motion of $\hvE^+(\vr,\omega)$ is derived as
\begin{align}& \label{eq:Maxwell} 
\rot\rot\hvE^+(\vr,\omega) -(\omega/c)^2\dieb(\vr,\omega) \hvE^+(\vr,\omega)
\nonumber \\ &
= \ii\muz\omega\hvJz(\vr,\omega) + \muz\omega^2\hvPex^+(\vr,\omega),
\end{align}
where $\dieb(\vr,\omega)$ is the background dielectric function,
$\mu_0$ is the magnetic permeability in vacuum,
and $\hvJz(\vr,\omega)$ governs the fluctuation of the electromagnetic fields
\cite{newQED}.

The third term on the right-hand side of Eq.~\eqref{eq:motion-exciton}
is the nonlinear term for the entangled-photon generation,
and it is approximately treated as follows.
First, we calculate the amplitude
$\beta_{\mu}(\omega) = \braket{\hex_{\mu}(\omega)}$ 
of excitons in linear regime from Eqs.~\eqref{eq:motion-exciton}
and \eqref{eq:Maxwell} by neglecting the nonlinear term
and by assuming a pump field as a homogeneous solution
of Eq.~\eqref{eq:Maxwell} in classical framework of light.
Then, the biexciton amplitude $\mathcal{B}_{\lambda}(\omega)$ 
is calculated from $\beta_{\mu}(\omega)$ and 
a phenomenologically introduced damping rate $\dampbx$ of biexcitons \cite{note}.
Considering a sufficiently strong pump power
compared to the vacuum fluctuation,
we can approximately rewrite the nonlinear term in Eq.~\eqref{eq:motion-exciton}
as $\sum_{\lambda,\nu}
  (\hbar\wex_{\mu}+\hbar\wex_{\nu}-\hbar\varOmega_{\lambda})
  \wvfbx_{\lambda,\mu,\nu} \int_{-\infty}^{\infty}\dd\omega'\
  \oexd_{\nu}(\omega-\omega') \mathcal{B}_{\lambda}(\omega')$,
where $\varOmega_{\lambda}$ is the eigen frequency of biexcitons 
in state $\lambda$
and $\wvfbx_{\lambda,\mu,\nu}$ is the coefficient giving
a biexciton state
$\ket{\lambda} 
= \half \sum_{\mu,\nu} \wvfbx_{\lambda,\mu,\nu} \oexd_{\mu} \oexd_{\nu}\ket{0}$.
While $\varOmega_{\lambda}$ and $\wvfbx_{\lambda,\mu,\nu}$
should be determined from Eq.~\eqref{eq:oHex},
we instead assume them as follows.
If the center-of-mass motion of the lowest biexciton level (zero angular momentum)
is confined in a crystal sufficiently larger than its Bohr radius,
we can approximately express the coefficient as
$\wvfbx_{\lambda,\mu,\nu} = \delta_{\lambda,\mu,\nu} \varPhi
  \int\dd\vr\ G_{\lambda}(\vr) g_{\mu}^*(\vr) g_{\nu}^*(\vr)$,
where $|\varPhi|^2$ represents the effective volume of relative motion of biexciton,
$\delta_{\lambda,\mu,\nu}$ describes the polarization selection rule
in the excitation and collapse of biexcitons,
and $G_{\lambda}(\vr)$ is the center-of-mass wavefunction of biexciton state $\lambda$.

Under the above approximation,
by simultaneously solving Eqs.~\eqref{eq:motion-exciton} 
and \eqref{eq:Maxwell} up to the lowest order, 
we evaluate correlation functions of $\hvE^+(\vr,\omega)$
from commutation relations of fluctuation operators $\hat{\cal D}_{\mu}(\omega)$
and $\hvJz(\vr,\omega)$ \cite{newQED}.
The scattering intensity is evaluated by the first-order
correlation $C^{(1)}(x) = \braket{\hvE^-(x)\hvE^+(x)}$
at $x = (\vr, \omega)$, where $\vr$ is a position outside the material
and the scattering frequency $\omega$ differs
from the pump frequency $\win$.
The two-photon coincidence intensity is evaluated by the second-order
correlation $C^{(2)}(x_1,x_2) = \braket{\hvE^-(x_1)\hvE^-(x_2)\hvE^+(x_2)\hvE^+(x_1)}
= C^{(2)}_{S}(x_1,x_2) + C^{(2)}_{N}(x_1,x_2)$.
Here, $C^{(2)}_{S}(x_1,x_2)$ is the intensity of correlated photons
satisfying the frequency conservation
$2\win = \omega_1 + \omega_2$,
while $C^{(2)}_{N}(x_1,x_2) = C^{(1)}(x_1)C^{(1)}(x_2)$
is the intensity of accidental pairs
and is written as a product of two first-order correlations \cite{note4}.

We consider a CuCl film with thickness $\thick$,
and the $z$-axis is perpendicular to its surface.
We consider only the lowest relative motion of excitons and biexcitons 
and assume sinusoidal wavefunctions for their
center-of-mass motion (indexed by $n$) as seen in Fig.~\ref{fig:1}(b).
We assume their standard parabolic energy dispersions
and follow Ref.~\onlinecite{Crossover} for other excitonic parameters.
Concerning biexcitons, its translational mass is $\mbx = 2.3\mex$, 
and the binding energy is $\Dbx = 32.2\;\meV$ \cite{Biexciton}.
From Ref.~\onlinecite{Akiyama}, we consider
the effective volume as $|\rmfbx|^2 = 80\;\nm^3$
and the damping rate as $\dampbx = 13.2\;\microeV$.
The pumping light is a continuous plane wave perpendicular to the film,
and has the same frequency as the giant two-photon absorption of CuCl
except in Fig.~2(b), where $\win$ is tuned to resonantly excite
$n=6$ biexciton state.
In both cases, $\hbar\win \simeq \hbar\wT - \Dbx/2$,
where $\hbar\wT = 3.2022\;\eV$ is the transverse bare exciton energy.
We consider the pump power as $I = 10~\mu\mathrm{W}$ (except in Fig.~\ref{fig:3}(b))
as used in Ref.~\onlinecite{Oohata},
while spectral shapes of the RHPS are not modified by changing $I$.
The scattering angle $\theta$ is defined as providing
the in-plane wavenumber $k_{\parallel} = (\wT/c) \sin\theta$,
and the above correlation functions are transformed into 
$k_{\parallel}$-space as $x = (\theta,z,\omega)$,
while they depend on $z$ only for whether the scattering fields
are forward or backward with respect to the pump propagation.
We define the horizontal (H) and vertical (V) directions of the polarization
with respect to the scattering plane.

\begin{figure}[tbp] 
\includegraphics[width=\hsize]{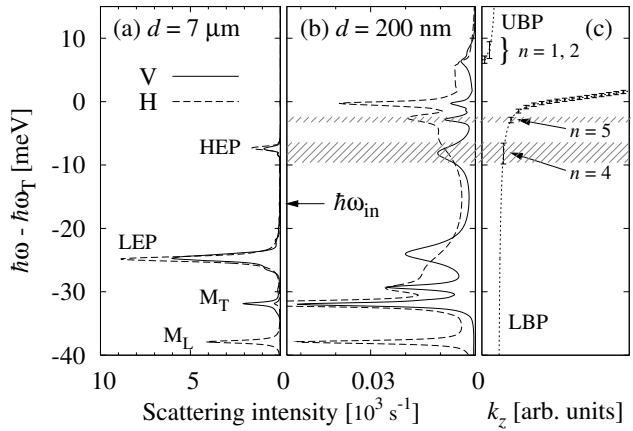}
\caption{Forward RHPS spectra of CuCl film 
with thicknesses of 
(a) $d = 7\;\mu\mathrm{m}$ and (b) $d = 200\;\mathrm{nm}$.
V- and H-polarized spectra are plotted by solid and dashed lines, respectively.
(c) Dotted lines represent the dispersion curves of upper and lower branch polaritons (UBP and LBP), and vertical bars indicate the resonance frequency of V-polarized exciton-photon coupled modes in the film with $d = 200\;\mathrm{nm}$.
The length of the bars represents
the sum of radiative and nonradiative
decay rates of the coupled modes.
$n$ denotes the index of the original exciton state of each coupled mode.
Parameters:  $\gamma_{\mathrm{ex}} = 0.5~\mathrm{meV}$, $\theta = 60^{\circ}$.}
\label{fig:epsart}
\end{figure}
Fig.~\ref{fig:epsart} shows the polarization-resolved spectra
of forward scattering intensity (proportional to $C^{(1)}$ \cite{note2})
at thicknesses of (a) $7\;\mu\mathrm{m}$ and (b) 200\;nm.
The nonradiative damping rate of excitons is 
$\gamma_{\mathrm{ex}} = 0.5~\mathrm{meV}$
and the scattering angle is $\theta = 60^{\circ}$.
In the case of the bulk-like thickness [Fig.~\ref{fig:epsart}(a)], 
the four peaks, namely, $\text{M}_{\text{L}}$, 
$\text{M}_{\text{T}}$, LEP, and HEP (lower and higher energy polariton)
are reproduced
as observed in experiments \cite{Biexciton,Eda,Oohata}.
The LEP and HEP correspond to the RHPS process, 
while $\mathrm{M_T}$ and $\mathrm{M_L}$ are caused by the biexciton
relaxation to the transverse and longitudinal excitons, respectively.
The frequencies of LEP and HEP depend on
scattering angle under the conservation of energy and wavevector
in bulk materials.
Decreasing the film thickness,
the LEP and HEP peaks diminish due to the relax
of the wavevector conservation.
Instead, multiple peaks appear in the LEP-HEP frequency region
as seen in Fig.~\ref{fig:epsart}(b).
These anomalous peaks can be interpreted by the energy level structure
of exciton-photon coupled modes 
in the nano film \cite{Shouji,IKS,Crossover,Ichimiya},
which are shown in Fig.~\ref{fig:epsart}(c) as vertical bars
with length of sum of the radiative and nonradiative damping rates.
The scattering spectra
reflect the anomalous level structure of the coupled modes,
because a biexciton relaxes into a coupled mode by 
emitting a photon (a peak on the lower energy side)
with satisfying the energy conservation.
It is worth noting that we can significantly modify
resonance frequencies and radiative lifetimes of
the exciton-photon coupled modes by designing the material structure,
i.e., material shape, size, arrangement, and external environments
\cite{Crossover,Ichimiya,Shouji}.
Since the anomalous level structures depend even on propagation angle and polarization direction, the nano-structured materials
have much degrees of freedom to control the frequencies, angles, polarizations 
and phase difference of generated entangled state.
This aspect is essentially different from the cases of bulk crystals \cite{Kwiat, Eda, Oohata} and 
of the single quantum dot systems \cite{Akopian,Stevenson}.

For the high-power generation of the entangled photons,
the important factors are the generation efficiency and 
also the statistical accuracy, i.e., the amount of accidental pairs.
For a pumping beam with power $I$,
the signal intensity $S \propto C^{(2)}_S$ (amount of correlated pairs) 
is proportional to $I^{2}$, 
while the noise intensity $N \propto C^{(2)}_N$ (amount of accidental pairs)
is proportional to $I^{4}$,
because an accidental pair is involved with two biexcitons.
This implies that, by increasing the pump power $I$,
the $S/N$ ratio decreases in contrast to the increase of $S$ \cite{Oohata}.
Here, we introduce another measure termed ``performance'' $P$
defined as signal intensity $S$
under a certain $S/N$ ratio
($I$ is tuned to give this ratio).
This quantity does not depend on $I$,
and represents the material potential for generating strong
and qualified entangled-photon beams.
While we suppose $S/N = 20$ as reported in Ref.~\onlinecite{Oohata}
to calculate $P = S^2/20N$,
the spectral shape of the performance 
does not depend on a chosen $S/N$ ratio.

\begin{figure}[tb] 
\includegraphics[width=\hsize]{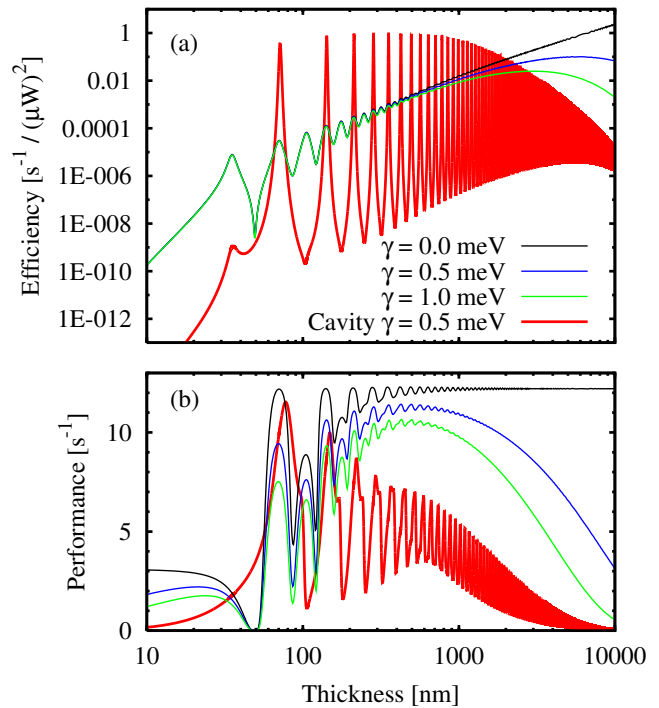}
\caption{Thickness dependence of (a) generation efficiency $S/I^2$ 
and (b) performance $P$.
Dielectric constant of outside is $\dieb$, 
scattering frequencies are the same as that of pumping field,
and scattering is forward with $\theta=0^{\circ}$ (black, blue, and green).
The red lines represent backward emission from a CuCl film with an optical cavity, whose mode frequency is tuned to $\wT$.}
\label{fig:3}
\end{figure}
Fig.~\ref{fig:3} shows thickness dependences of (a) generation efficiency
$S/I^2$ and (b) performance $P$.
For simplicity, we suppose that
the two scattering fields are perpendicular to the film ($\theta = 0^{\circ}$)
and frequencies are $\omega_{1/2} = \win\pm 0^+$.
The black, blue, and green lines ($\dampex = 0$, 0.5, and $1.0\;\mathrm{meV}$,
respectively) represent the forward emission
from a CuCl film existing in a dielectric medium with $\dieb$.
As seen in Fig.~\ref{fig:3}(a) and also in 
Ref.~\onlinecite{Savasta}, the thickness maximizing the generation efficiency 
is changed by $\dampex$, 
and it is in the order of micrometers or more for CuCl crystals.

However, as seen in Fig.~\ref{fig:3}(b),
the performance decreases with increasing thickness for non-zero $\dampex$
at thickness range of micrometers,
because the amplitudes of scattering fields decrease
during the propagation in the absorptive film (see Fig.~\ref{fig:1}(a)),
and the performance does not increase even if $\dampex = 0$.
The maximum performance in Fig.~\ref{fig:3}(b) is the ideal quantity,
and it only depends on measurement conditions, 
such as resolutions of angle and frequency, but not on material parameters
\cite{note2}.
Therefore, the generation efficiency
(generation probability for one pump pulse)
is limited by the statistical accuracy ($S/N$ ratio)
when we use bulk crystals \cite{Oohata}.
However, at thickness of from 50\;nm to 1000\;nm,
it is worth noting that
nearly ideal performance values can be obtained at particular thicknesses
even if $\dampex$ is non-zero.
This is because the radiative decay is dominant in such nano films
owing to the exciton superradiance \cite{Ichimiya,Crossover},
and the entangled pairs can go outside the film
without decreasing the amplitude.
Therefore, nano films generally show a high performance,
and this rapid decay is also desired
for the high repetition excitation,
which also recovers the signal intensity with keeping the $S/N$ ratio \cite{Oohata}.
A good performance is obtained only when
the resonance energy of exciton-photon coupled mode is
just equal to a half of the biexciton energy.
This condition appears as an oscillatory behavior in this crossover thickness region, because the center-of-mass of
biexcitons is confined and the
energy levels of coupled modes are modified by changing
the thickness.

However, as seen in Fig.~\ref{fig:3}(a), 
the generation efficiency of nano films
is much lower than that of bulk crystals.
In other words, 
a strong pump power is required to achieve a sufficient signal intensity
at the thickness range of nanometers.
This problem can be overcome
by controlling the light-matter coupling
by using an optical cavity.
Avoiding essential modifications of the biexciton level scheme
and considering an existing sample, we treat a cavity
with a low quality factor (Q-factor) as reported in the experiment
\cite{Cavity},
namely a CuCl film in a distributed Bragg reflector (DBR) cavity composed 
by $\mathrm{PbF_2}$ and $\mathrm{PbBr_2}$.
The DBR reflectors are considered by the background dielectric function
$\dieb(\vr,\omega)$.
The red lines in Fig.~\ref{fig:3} represent the backward emission
from the cavity structure,
where the cavity mode frequency is tuned to $\wT$,
$\dampex = 0.5\;\mathrm{meV}$,
and the periods of the incident- and transmitted-side are 4 and 16,
respectively.
This system corresponds to the weak bipolariton regime,
where the energy splitting between polariton and biexciton levels are small
compared to their broadening. 
This situation is in contrast with 
Ref.~\onlinecite{Ajiki}, where the strong enhancement of 
entangled-photon generation from a quantum well in a cavity 
with higher Q-factor has been discussed 
based on the biexcitonic cavity-QED picture or the strong bipolariton picture.
As seen in Fig.~\ref{fig:3}(a),
since the biexcitons are effectively created,
the generation efficiency is significantly enhanced at thickness of nanometers,
and it is larger than the maximum value
in the previous situation ($\dampex = 0.5\;\mathrm{meV}$, blue line).
The enhancement also occurs when the energy of polariton 
(exciton-photon coupled mode) is equal to a half of biexciton energy,
which is consistent with Ref.~\onlinecite{Ajiki}. 
Comparing to the previous data (blue line),
the period of the oscillation is doubled,
because the RHPS process involving biexcitons 
with odd-parity center-of-mass motion
is forbidden in an one-sided optical cavity.
On the other hand, as seen in Fig.~\ref{fig:3}(b),
at thickness of micrometers,
the performance is suppressed compared to that by the bare CuCl film.
This is because of the multiple reflection inside the cavity,
and the scattered fields nonradiatively decay during the propagation.
In contrast, at thickness of nanometers, especially at 80\;nm,
the performance almost keeps the ideal quantity,
because of the enhancement of excitons' radiative decay rate
by the optical cavity
or the exciton superradiance \cite{Ichimiya,Crossover}.
In this way, for the optimum condition, we can highly control both 
the generation efficiency and the performance 
by manipulating the confinement of biexcitons and 
the level structure of the exciton-photon coupled modes
in a nano film embedded in an optical cavity.

In summary, 
semiconductor nano-films have much degrees of freedom to control
generated states of entangled photon pairs.
They show a high performance from the viewpoint 
of statistical accuracy,
and the trade-off problem in bulk materials can be overcome
by the enhancement of radiative decay of excitons.
By using a cavity under the weak bipolariton regime,
the generation efficiency can be enhanced
with keeping the high performance.
We believe that our results make a breakthrough
of entangled-photon generation with both a high generation efficiency
and the ideal statistical accuracy.

The authors thank K.~Edamatsu, G.~Oohata, H.~Ajiki, and H.~Oka
for their helpful discussions. 
This work was partially supported by the Japan Society 
for the Promotion of Science (JSPS); 
a Grant-in-Aid for Creative Science Research, 17GS1204, 2005; 
and a Grant-in-Aid for JSPS Research Fellows.

\def\theequation{A.\arabic{equation}}
\setcounter{equation}{0}
\section{Supplementary material}
\subsection{Hamiltonian}
We consider the Hamiltonian on the main paper as
\begin{equation}
\oH = \oHex + \oHres + \oHint + \oHem,
\end{equation}
where $\oHex$ describes the excitonic systems,
$\oHres$ represents a reservoir for the nonradiative damping of excitons,
$\oHint$ is the exciton-photon interaction,
and $\oHem$ describes the electromagnetic fields and a background dielectric
medium,
which is the Hamiltonian discussed by Suttorp et al.~in 
Ref.~\cite{suttorp04} and also used in Ref.~\cite{newQED}.
We consider the Hamiltonian of excitonic system as
\begin{equation} \label{eq:def-Hamilt-ex} 
\oHex = \sum_{\mu} \hbar\wex_{\mu} \oexd_{\mu} \oex_{\mu}
+ \half \sum_{\mu,\mu',\nu,\nu'} V_{\mu,\nu;\mu',\nu'}
  \oexd_{\mu} \oexd_{\nu} \oex_{\nu'} \oex_{\mu'},
\end{equation}
where $\oex_{\mu}$ is the annihilation operator of an exciton
in state $\mu$ and $\wex_{\mu}$ is its eigenfrequency.
We treat the excitons as pure bosons satisfying
\begin{subequations}
\begin{align}
\left[ \oex_{\mu}, \oexd_{\mu'} \right] & = \delta_{\mu,\mu'}, \\
\left[ \oex_{\mu}, \oex_{\mu'} \right] & = 0,
\end{align}
\end{subequations}
and their non-bosonic behavior is described 
by the exciton-exciton interaction, the second term
in Eq.~\eqref{eq:def-Hamilt-ex}.
The reservoir $\oHres$ is written as
\begin{align}
\oHres & = \sum_{\mu} \int_0^{\infty}\dd\wres\ \bigl\{
    \hbar\wres \oresd_{\mu}(\wres) \ores_{\mu}(\wres)
\nonumber \\ & \quad
+ \left[ \oex_{\mu} + \oexd_{\mu} \right]
  \left[ \ccxr_{\mu}(\wres) \ores_{\mu}(\wres)
       + \ccxr^*_{\mu}(\wres) \oresd_{\mu}(\wres) \right] \bigr\},
\end{align}
where $\ores_{\mu}(\wres)$ is the annihilation operator of harmonic
oscillator with frequency $\wres$
interacting with excitons in state $\mu$,
and $\ccxr_{\mu}(\wres)$ is the coupling coefficient.
The oscillators are independent with each other
and satisfy the following commutation relations:
\begin{subequations} \label{eq:[ores,ores]} 
\begin{align}
[ \ores_{\mu}(\wres), \oresd_{\mu'}(\wres') ]
& = \delta_{\mu,\mu'} \delta(\wres-\wres'), \\
[ \ores_{\mu}(\wres), \ores_{\mu'}(\wres') ]
& = 0.
\end{align}
\end{subequations}
Further, $\oHint$ is simply written as
a product of electric field $\ovE(\vr)$ and excitonic polarization
$\ovPex(\vr)$:
\begin{equation}
\oHint = - \int\dd\vr\ \ovPex(\vr) \cdot \ovE(\vr).
\end{equation}
Here, the excitonic polarization is represented as
\begin{equation}
\ovPex(\vr) = \sum_{\mu} \vdimP_{\mu}(\vr) \oex_{\mu} + \Hc,
\end{equation}
where the coefficient $\vdimP_{\mu}(\vr)$ is represented by the exciton
center-of-mass wavefunction $\cmfex_{\mu}(\vr)$ and 
unit vector $\vunit_{\mu}$ of polarization direction as
\begin{equation}
\vdimP_{\mu}(\vr) = \dimP \vunit_{\mu} \cmfex_{\mu}(\vr).
\end{equation}
The absolute value of $\dimP$ can be estimated 
by the longitudinal-transverse (LT) splitting energy
$\DLT = |\dimP|^2/\diez\dieb$ of excitons
and by background dielectric constant $\dieb$ of excitonic medium.

\subsection{Equation of motion}
According to Ref.~\cite{newQED}
or the QED theories of dispersive and absorbing media
\cite{huttner92,knoll01,suttorp04},
the equation of motion of the electric field is derived 
in frequency domain as
\begin{align}& \label{eq:Maxwell-E-Jz-Pex} 
\rot\rot\hvE^+(\vr,\omega)
- \frac{\omega^2}{c^2}\dieb(\vr,\omega)\hvE^+(\vr,\omega)
\nonumber \\ &
= \ii\muz\omega\hvJz(\vr,\omega) + \muz\omega^2\hvPex^+(\vr,\omega).
\end{align}
Here, $\dieb(\vr,\omega)$ is the dielectric function
of the background medium.
$\hvJz(\vr,\omega)$ describes the fluctuation
of electromagnetic fields and satisfies
\begin{align} \label{eq:[hvJz,hvJz]} 
&   \left[ \hvJz(\vr,\omega), \dgg{\hvJz(\vr',{\omega'}^*)} \right]
= \left[ \hvJz(\vr,\omega), \hvJz(\vr',-\omega') \right]
\nonumber \\ &
= \delta(\omega-\omega') \delta(\vr-\vr') \frac{\diez\hbar\omega^2}{\pi}
    \Im[\dieb(\vr,\omega)] \munit.
\end{align}

In the same manner as in Ref.~\cite{newQED},
under the rotating-wave approximation (RWA),
we obtain the equation of excitons' motion in frequency domain as
\begin{align}& \label{eq:motion-hex} 
\left[ \hbar\wex_{\mu}-\hbar\omega-\ii\dampex/2 \right]
\hex_{\mu}(\omega)
\nonumber \\ &
= \int\dd\vr\ \vdimP_{\mu}^*(\vr) \cdot \hvE^+(\vr,\omega)
  + \hrsrc_{\mu}(\omega)
\nonumber \\ & \quad
- \sum_{\nu} \sum_{\mu',\nu'} V_{\mu,\nu;\mu',\nu'}
  \int_{-\infty}^{\infty}\dd t\ \frac{\ee^{\ii\omega t}}{2\pi}
  \oexd_{\nu}(t) \oex_{\nu'}(t) \oex_{\mu'}(t),
\end{align}
where $\dampex$ is the nonradiative damping width
(defined by $\{\ccxr_{\mu}(\wres)\}$ in Eq.~(D7) of Ref.~\cite{newQED}),
and $\hrsrc_{\mu}(\omega)$ is the fluctuation operator satisfying
\begin{align}& \label{eq:[hrsrc,hrsrc]} 
\left[ \hrsrc_{\mu}(\omega), \dgg{\hrsrc_{\mu'}({\omega'}^*)} \right]
= \left[ \hrsrc_{\mu}(\omega), \hrsrc_{\mu'}(-\omega') \right]
\nonumber \\ &
= \delta_{\mu,\mu'} \delta(\omega-\omega') \frac{\hbar\dampex}{2\pi}.
\end{align}
The last term on the right hand side of Eq.~\eqref{eq:motion-hex}
is the nonlinear term due to the exciton-exciton interaction.

Here, we define a new operator
\begin{equation} \label{eq:def-obx} 
\obx_{\lambda}
\equiv \half \sum_{\mu,\nu} \wvfbx_{\lambda,\mu,\nu}^* \oex_{\nu} \oex_{\mu},
\end{equation}
which annihilates a biexciton (two excitons) in state $\lambda$
or describes a two-exciton eigen state $\obxd_{\lambda}\ketg$
by applying it to the matter ground state $\ketg$.
The coefficient $\wvfbx_{\lambda,\mu,\nu}$ is invariant 
by the exchange of two exciton indices as
\begin{equation} \label{eq:invariant-mu-nu} 
\wvfbx_{\lambda,\mu,\nu} = \wvfbx_{\lambda,\nu,\mu}.
\end{equation}
Further, it is ortho-normal as
\begin{equation}
\half\sum_{\mu,\nu} \wvfbx_{\lambda,\mu,\nu} \wvfbx_{\lambda',\mu,\nu}^*
= \delta_{\lambda,\lambda'},
\end{equation}
and also has a completeness
\begin{equation} \label{eq:complete-wvfbx} 
\sum_{\lambda} \wvfbx_{\lambda,\mu,\nu}
\wvfbx_{\lambda,\mu',\nu'}^*
= \delta_{\mu,\mu'} \delta_{\nu,\nu'}
+ \delta_{\mu,\nu'} \delta_{\nu,\mu'}.
\end{equation}
From the excitonic Hamiltonian \eqref{eq:def-Hamilt-ex},
the coefficient $\wvfbx_{\lambda,\mu,\nu}$ 
and eigen frequency $\wbx_{\lambda}$
of biexciton eigen state $\lambda$
should satisfy
\begin{equation} \label{eq:wvfbx-satisfy-1} 
( \hbar\omega_{\mu} + \hbar\omega_{\nu} ) \wvfbx_{\lambda,\mu,\nu}
+ \sum_{\mu',\nu'} V_{\mu,\nu;\mu',\nu'} \wvfbx_{\lambda,\mu',\nu'}
= \hbar\wbx_{\lambda} \wvfbx_{\lambda,\mu,\nu}.
\end{equation}
By using Eqs.~\eqref{eq:invariant-mu-nu} and \eqref{eq:complete-wvfbx},
we can rewrite Eq.~\eqref{eq:def-obx} as
\begin{equation}
\sum_{\lambda} \wvfbx_{\lambda,\mu,\nu} \obx_{\lambda}
= \oex_{\nu} \oex_{\mu}.
\end{equation}
Therefore, from this relation and Eq.~\eqref{eq:wvfbx-satisfy-1},
we can rewrite Eq.~\eqref{eq:motion-hex} as
\begin{align}& \label{eq:motion-hex-2} 
\left[ \hbar\wex_{\mu}-\hbar\omega-\ii\dampex/2 \right]
\hex_{\mu}(\omega)
\nonumber \\ &
= \int\dd\vr\ \vdimP_{\mu}^*(\vr) \cdot \hvE^+(\vr,\omega)
  + \hrsrc_{\mu}(\omega)
\nonumber \\ & \quad
+ \sum_{\lambda,\nu}( \hbar\wex_{\mu}+ \hbar\wex_{\nu}
                     - \hbar\wbx_{\lambda})
  \wvfbx_{\lambda,\mu,\nu}
\nonumber \\ & \quad \times
  \int_{-\infty}^{\infty}\dd\omega'\
  \dgg{\hex_{\nu}(\omega'-\omega)} \hbx_{\lambda}(\omega').
\end{align}
On the other hand,
the equation of motion of $\obx_{\lambda}$ is derived
in frequency domain as
\begin{align}& \label{eq:motion-hbx} 
(\hbar\wbx_{\lambda}-\hbar\omega) \hbx_{\lambda}(\omega)
\nonumber \\ &
= \sum_{\mu,\nu}\wvfbx_{\lambda,\mu,\nu}^*\ \int_{-\infty}^{\infty}\dd\omega'\
    (\hbar\wex_{\nu}-\hbar\omega') \hex_{\nu}(\omega')
    \hex_{\mu}(\omega-\omega').
\end{align}
In principle, the biexciton-resonant hyperparametric scattering (RHPS) process 
is described by equations of motion
\eqref{eq:Maxwell-E-Jz-Pex}, \eqref{eq:motion-hex-2}, 
and \eqref{eq:motion-hbx},
and commutation relations \eqref{eq:[hvJz,hvJz]} and \eqref{eq:[hrsrc,hrsrc]}.
However, in the actual calculation, we use the following approximation.

\subsection{Approximation for RHPS process}
We suppose that a coherent light beam resonantly excites biexcitons
and their amplitude is large enough compared to the vacuum fluctuation.
In this case, if we do not consider the other higher processes,
the biexciton operator in the nonlinear term of Eq.~\eqref{eq:motion-hex-2}
can be replaced by its amplitude
$\mathcal{B}_{\lambda}(\omega') = \braket{\hbx_{\lambda}(\omega')}$.
Further, we replace $\hex_{\nu}(\omega'-\omega)$
in the nonlinear term by $\hexone_{\nu}(\omega'-\omega)$,
which satisfies the linear equation
\begin{align}& \label{eq:motion-hexone} 
\left[ \hbar\wex_{\mu}-\hbar\omega-\ii\dampex/2 \right]
\hexone_{\mu}(\omega)
\nonumber \\ &
= \int\dd\vr\ \vdimP_{\mu}^*(\vr) \cdot \hvE^+(\vr,\omega)
  + \hrsrc_{\mu}(\omega).
\end{align}
Simultaneously solving this equation and Eq.~\eqref{eq:Maxwell-E-Jz-Pex},
$\hexone_{\mu}(\omega)$ can be expressed by
the fluctuation operators $\hvJz(\vr,\omega)$ and $\hrsrc_{\mu}(\omega)$.
Under the above approximation, we simultaneously solve
Eq.~\eqref{eq:Maxwell-E-Jz-Pex} and
\begin{align}& \label{eq:motion-hex-3} 
\left[ \hbar\wex_{\mu}-\hbar\omega-\ii\dampex/2 \right]
\hex_{\mu}(\omega)
\nonumber \\ &
\simeq \int\dd\vr\ \vdimP_{\mu}^*(\vr) \cdot \hvE^+(\vr,\omega)
  + \hrsrc_{\mu}(\omega)
\nonumber \\ & \quad
+ \sum_{\lambda,\nu}( \hbar\wex_{\mu}+ \hbar\wex_{\nu}
                     - \hbar\wbx_{\lambda})
  \wvfbx_{\lambda,\mu,\nu}
\nonumber \\ & \quad \times
  \int_{-\infty}^{\infty}\dd\omega'\
  \dgg{\hexone_{\nu}(\omega'-\omega)} \mathcal{B}_{\lambda}(\omega'),
\end{align}
and then represent $\hvE^+(\vr,\omega)$
by the fluctuation operators $\hvJz(\vr,\omega)$ and $\hrsrc_{\mu}(\omega)$.
The calculation procedure is straightforward
by using the Green's function technique
(see section \ref{sec:Green} or Ref.~\cite{newQED}).

For the calculation of $\mathcal{B}_{\lambda}(\omega)$,
we suppose that the biexciton amplitude does not
decrease by the RHPS process,
because its contribution is small compared to the incident light.
Under this approximation,
by phenomenologically introducing a damping constant $\dampbx$,
the biexciton amplitude is obtained from Eq.~\eqref{eq:motion-hbx} as
\begin{align}& \label{eq:<hbx>=<hexone>2} 
\braket{\hbx_{\lambda}(\omega)}
\simeq \frac{1}{\hbar\wbx_{\lambda}-\hbar\omega-\ii\dampbx/2}
  \sum_{\mu,\nu}\wvfbx_{\lambda,\mu,\nu}^*\
\nonumber \\ & \quad \times
  \int_{-\infty}^{\infty}\dd\omega'\
    (\hbar\wex_{\nu}-\hbar\omega') \braket{\hexone_{\nu}(\omega')}
    \braket{\hexone_{\mu}(\omega-\omega')},
\end{align}
where $\braket{\hexone_{\nu}(\omega')}$ can be calculated
from Eqs.~\eqref{eq:Maxwell-E-Jz-Pex} and \eqref{eq:motion-hexone}
by considering an incident light beam
as a homogeneous solution of Eq.~\eqref{eq:Maxwell-E-Jz-Pex}.
Under the weak bipolariton regime,
where the coupling between polariton and biexciton
is small enough compared to their broadening,
the approximated expression \eqref{eq:<hbx>=<hexone>2}
of biexciton amplitude is sufficient for the discussion of RHPS process.

\subsection{Model of biexcitons\label{sec:wvfbx}}
Although $\wvfbx_{\lambda,\mu,\nu}$ and $\wbx_{\lambda}$ should be determined 
from \eqref{eq:wvfbx-satisfy-1} for given nonlinear coefficient $V_{\mu,\nu;\mu',\nu'}$
in principle, we instead suppose them from the experimental results.
This treatment is useful because we already know many parameters
of the lowest biexciton level in CuCl
by the longstanding experimental and theoretical studies \cite{Biexciton}.

It is well known that the lowest level of biexciton in CuCl
is singlet and has zero angular momentum
because of the exchange interaction 
between two electrons and between two holes \cite{Biexciton}.
Since we suppose the resonant excitation of the lowest level,
we only consider the lowest relative motion of biexciton.
Further, according to the RHPS experiments in Ref.~\onlinecite{tokunaga99},
the lowest biexciton state mainly consists of $1s$ excitons,
and the contribution from the higher exciton states was estimated
in the order of $10^{-4}$.
Therefore, we consider only $1s$ relative motion of excitons,
which has a degree of freedom of polarization direction $\xi_{\mu}=\{x,y,z\}$.
The coefficient $\wvfbx_{\lambda,\mu,\nu}$
is proportional to the polarization selection rule as
\begin{equation}
\delta_{\lambda,\mu,\nu} = \delta_{\xi_{\mu},\xi_{\nu}}.
\end{equation}
Considering the relative motion $\varPsi(\vr)$ of two excitons
in the lowest biexciton level, the coefficient is written as
\begin{equation}
\wvfbx_{\lambda,\mu,\nu} = \delta_{\lambda,\mu,\nu}
\int\dd\vr\int\dd\vr'\ \varPsi(\vr') \cmfbx_{\lambda}(\vr)
  \cmfex_{\mu}^*(\vr+\vr') \cmfex_{\nu}^*(\vr),
\end{equation}
where $\cmfex_m(\vr)$ and $\cmfbx_l(\vr)$ are 
center-of-mass wavefunctions of excitons and biexcitons,
respectively.
Here, we suppose that the Bohr radius of biexciton
is much smaller than the crystal size,
and then we approximate the above expression as
\begin{equation} \label{eq:mwvfbx} 
\wvfbx_{\lambda} \simeq \delta_{\lambda,\mu,\nu}
\rmfbx \int\dd\vr\ \cmfbx_{\lambda}(\vr)\
  \cmfex_{\mu}^*(\vr)\ \cmfex_{\nu}^*(\vr),
\end{equation}
where $\rmfbx$ is defined as
\begin{equation}
\rmfbx \equiv \int\dd\vr\ \varPsi(\vr),
\end{equation}
and $|\rmfbx|^2$ represents 
the effective volume of the lowest biexciton state.
Therefore, we can separately consider 
the relative and center-of-mass motions of biexcitons.
$|\rmfbx|^2$ has been estimated by an experiment \cite{Akiyama},
and was also used as a parameter in calculation \cite{matsuura95}.

\subsection{Observables\label{sec:observe}}
For the one-photon scattering intensity at $\vr$
with polarization direction $\vunit$ and frequency $\omega$
(not equal to pump frequency $\win$),
its intensity is proportional to
the first-order correlation function
\begin{equation} \label{eq:intone} 
C^{(1)}(\vr,\vunit,\omega)
= \vunit \cdot \braket{ \hvE^-(\vr,\omega) \hvE^+(\vr,\omega) }
\cdot \vunit.
\end{equation}
This function can be calculated by the commutation relations
\eqref{eq:[hvJz,hvJz]} and \eqref{eq:[hrsrc,hrsrc]}.
The two-photon coincidence intensity
between $(\vr_1, \vunit_1, \omega_1)$
and $(\vr_2, \vunit_2, \omega_2)$ is proportional to
the second-order correlation function
\begin{widetext}
\begin{align} \label{eq:inttwo} 
C^{(2)}(\vr_1,\vunit_1,\omega_1;\vr_2,\vunit_2,\omega_2)
& = \braket{ \vunit_1 \cdot \hvE^-(\vr_1,\omega_1)
         \vunit_2 \cdot \hvE^-(\vr_2,\omega_2)
         \vunit_2 \cdot \hvE^+(\vr_2,\omega_2)
         \vunit_1 \cdot \hvE^+(\vr_1,\omega_1) }
\nonumber \\
& = C^{(2)}_S(\vr_1,\vunit_1,\omega_1;\vr_2,\vunit_2,\omega_2)
+ C^{(2)}_N(\vr_1,\vunit_1,\omega_1;\vr_2,\vunit_2,\omega_2),
\end{align}
\end{widetext}
where we neglect the interference term that appears only under $\omega_1 = \omega_2$.
$C^{(2)}_S$ represents the signal intensity
or the number of correlated photon pairs,
which satisfies the energy conservation $\omega_1 + \omega_2 = 2\win$.
On the other hand, 
$C^{(2)}_N$ appears for arbitrary pair of $\omega_1$ and $\omega_2$.
This represents the accidental coincidence of emitted photons
from independent biexcitons,
because it is just the product of two first-order correlation functions
\begin{align}
& C^{(2)}_N(\vr_1,\vunit_1,\omega_1;\vr_2,\vunit_2,\omega_2)
\nonumber \\ &
\equiv C^{(1)}(\vr_1,\vunit_1,\omega_1) C^{(1)}(\vr_2,\vunit_2,\omega_2).
\end{align}

\subsection{Solving wave equation} \label{sec:Green}
Here, we explain how we simultaneously solve Eq.~\eqref{eq:Maxwell-E-Jz-Pex}
and equation of exciton motion.
By using the dyadic Green's function satisfying
\begin{equation} \label{eq:satisfied-mG} 
\rot\rot\mG(\vr,\vr',\omega)
- \frac{\omega^2}{c^2}\dieb(\vr,\omega)\mG(\vr,\vr',\omega)
= \delta(\vr-\vr')\munit,
\end{equation}
we can rewrite Eq.~\eqref{eq:Maxwell-E-Jz-Pex} as
\begin{equation} \label{eq:E=Ez+G*Pex} 
\hvE^+(\vr,\omega) = \hvEz^+(\vr,\omega)
  + \muz\omega^2 \int\dd\vr'\ \mG(\vr,\vr',\omega) \cdot \hvPex^+(\vr',\omega),
\end{equation}
where $\hvEz^+(\vr,\omega)$ is the electric field 
in the background ($\oHem$) system, and it is defined as
\begin{equation} \label{eq:def-hvEz-hvJz} 
\hvEz^{+}(\vr,\omega) \equiv \ii\muz\omega
  \int\dd\vr'\ \mG(\vr,\vr',\omega) \cdot \hvJz(\vr',\omega).
\end{equation}
Further, from Eq.~\eqref{eq:[hvJz,hvJz]}, $\hvEz^+(\vr,\omega)$ satisfies
\cite{newQED}
\begin{align}& \label{eq:[Ez(r,w),Ez(r,w)]} 
  \left[ \hvEz^+(\vr,\omega), \hvEz^-(\vr',\omega') \right]
= \left[ \hvEz^+(\vr,\omega), \hvEz^+(\vr',-\omega') \right] \nonumber \\ &
= \delta(\omega-\omega') \frac{\muz\hbar\omega^2}{\ii2\pi}
  [\mG(\vr,\vr',\omega)-\mG^*(\vr,\vr',\omega)].
\end{align}
The expression of $\mG(\vr,\vr',\omega)$ in planar system
(dielectric multi-layer) is already known \cite{chew95}.

Substituting Eq.~\eqref{eq:E=Ez+G*Pex} into Eq.~\eqref{eq:motion-hex-3},
we obtain the equation set for exciton operators as
\begin{align}& \label{eq:self-consist-nl} 
\sum_{\mu'} \css_{\mu,\mu'}(\omega) \hex_{\mu'}(\omega)
= \int\dd\vr\ \vdimP_{\mu}^*(\vr) \cdot \hvEz^+(\vr,\omega)
+ \hrsrc_{\mu}(\omega) \nonumber \\ & \quad
+ \sum_{\lambda,\nu}
  (\hbar\wex_{\mu}+\hbar\wex_{\nu}-\hbar\varOmega_{\lambda})
  \wvfbx_{\lambda,\mu,\nu}
\nonumber \\ & \quad \times
  \int_{-\infty}^{\infty}\dd\omega'\
  \dgg{\hexone_{\nu}(\omega'-\omega)} \mathcal{B}_{\lambda}(\omega'),
\end{align}
where the coefficient in the left-hand side is defined as
\begin{align} \label{eq:def-css} 
&\css_{\mu,\mu'}(\omega)
\equiv \left[ \hbar\wex_{\mu} - \hbar\omega
            - \ii\dampex/2 \right] \delta_{\mu,\mu'}
\nonumber \\ & \quad
- \muz\omega^2\int\dd\vr\int\dd\vr'\ \vdimP_{\mu}^*(\vr)
    \cdot \mG(\vr,\vr',\omega) \cdot \vdimP_{\mu'}(\vr').
\end{align}
The last term of Eq.~\eqref{eq:def-css} is the renormalization
due to the exciton-exciton interaction via the electromagnetic fields.
Further, the equation of operator $\hexone_{\mu}(\omega)$ in the linear regime
is also rewritten as
\begin{equation}
\sum_{\mu'} \css_{\mu,\mu'}(\omega) \hexone_{\mu'}(\omega)
= \int\dd\vr\ \vdimP_{\mu}^*(\vr) \cdot \hvEz^+(\vr,\omega)
+ \hrsrc_{\mu}(\omega).
\end{equation}
This simultaneous linear equation set is solved
by the inverse matrix $\mcrd(\omega) = [\mcss(\omega)]^{-1}$,
and the commutation relation of $\hexone_{\mu}(\omega)$ is derived
in Ref.~\cite{newQED} as
\begin{subequations} \label{eq:[hexone,hexone]} 
\begin{align}&
\left[ \hexone_{\mu}(\omega), \dgg{\hexone_{\mu'}({\omega'}^*)} \right]
\nonumber \\ & \quad
= \delta(\omega-\omega') \frac{\hbar}{\ii2\pi}
    \left[ \crd_{\mu,\mu'}(\omega) - \crd_{\mu',\mu}^*(\omega) \right], \\
& \left[ \hexone_{\mu}(\omega), \hexone_{\mu'}(-\omega') \right]
= 0.
\end{align}
\end{subequations}
Further, Eq.~\eqref{eq:self-consist-nl} is rewritten as
\begin{align} \label{eq:self-consist-app} 
\hex_{\mu}(\omega)
&\simeq \hexone_{\mu}(\omega)
+ \sum_{\mu',\lambda,\nu} \crd_{\mu,\mu'}(\omega)
  (\hbar\wex_{\mu'}+\hbar\wex_{\nu}-\hbar\varOmega_{\lambda})
\nonumber \\ & \quad \times
  \wvfbx_{\lambda,\mu',\nu}
  \int_{-\infty}^{\infty}\dd\omega'\
  \dgg{\hexone_{\nu}(\omega'-\omega)} \mathcal{B}_{\lambda}(\omega'),
\end{align}
and then the expression of electric field is obtained 
by substituting this into Eq.~\eqref{eq:E=Ez+G*Pex}.

\subsection{Intensity estimation}
We have estimated the scattering intensity,
signal $S$, and noise $N$ of RHPS process
from the experimental data in Ref.~\cite{Oohata}.
For a bulk crystal, the reported maximum count of entangled pairs
(sum of $HH$ and $VV$)
is 50 in 300 seconds and the width of the pulse is 2\;ns.
Since the time resolution is 300\;ps \cite{Eda},
the number of pairs in unit time is roughly estimated as
\begin{equation}
\frac{1}{2} \times 50 \times \frac{2\;\ns}{300\;\ps}
\times \frac{1}{300\;\sec} \simeq 0.5\;\sec^{-1}.
\end{equation}
The reported signal-to-noise ratio is $S/N = 20$,
and then the number of uncorrelated pairs 
is $0.025\;\sec^{-1}$.
The reported pump power is $I=10\;\mu\Watt$.
For the intensity estimation of our calculation results,
we have suppose that these experimental results
correspond to the our data in the case of CuCl film in vacuum
with $\dampex = 0.5\;\meV$ and thickness of $7\;\microm$,
which is the optimum thickness for the generation efficiency $S/I^2$.
On the main paper, we consider a CuCl film with dielectric medium
with $\dieb$ or the film with distributed Bragg reflectors.



\begin{thebibliography}{10}
\bibitem{note4}
While neigther the surface scattering nor the luminescence
are not considered in the calculation,
accidental pairs mainly come from independent biexcitons \cite{Oohata}.
\bibitem{note2} We estimate the absolute values of the scattering intensity,
generation efficiency, and performance from the experimental values
in Ref.~\onlinecite{Oohata} \cite{note}.
\end{thebibliography}
\end{document}